\documentclass[aps,prd,twocolumn,titlepage,superscriptaddress,nofootinbib]{revtex4-2}

\usepackage{graphicx}      
\usepackage{amssymb}       
\usepackage{amsmath}       
\usepackage{latexsym}      
\usepackage{cancel}        
\usepackage[normalem]{ulem} 
\usepackage{url}           
\usepackage{soul}          
\usepackage{verbatim}      
\usepackage{multirow}      
\usepackage{mathrsfs}      
\usepackage{float}         
\usepackage[dvipsnames]{xcolor} 
\usepackage{mathtools}     
\usepackage{slashed}       
\usepackage{physics}       
\usepackage{epstopdf}      
\usepackage{subfigure}     
\usepackage{bbold}         
\usepackage{wasysym}       
\usepackage{feynmp}        

	\usepackage[colorlinks=true, citecolor=blue, linkcolor=blue, urlcolor=blue]{hyperref}
	
	\newcommand{\beq}{\begin{eqnarray}}
		\newcommand{\eeq}{\end{eqnarray}}

	\makeatletter
	\newcommand{\myBig}{\bBigg@{1.75}}
	\makeatother
	\usepackage{orcidlink}
	\usepackage{etoolbox}
\begin{document}
		
		\title{
			Interior geometry of black holes as a probe of first-order phase transition }

		\author{Zi-Qiang Zhao\orcidlink{0009-0009-7859-3655}}
		\affiliation{Liaoning Key Laboratory of Cosmology and Astrophysics, College of Sciences, Northeastern University, Shenyang 110819, China}

		\author{Zhang-Yu Nie\orcidlink{0000-0001-7064-247X}}
		\email{niezy@kust.edu.cn}
		\affiliation{Center for Gravitation and Astrophysics, Kunming University of Science and Technology, Kunming 650500, China}
		
		\author{Shao-Wen Wei\orcidlink{0000-0003-0731-0610}}
		\email{weishw@lzu.edu.cn}
		\affiliation{Lanzhou Center for Theoretical Physics, Key Laboratory of Theoretical Physics of Gansu Province, Key Laboratory of Quantum Theory and Applications of MoE, Gansu Provincial Research Center for Basic Disciplines of Quantum Physics, Lanzhou University, Lanzhou 730000, China}
		
		
		\author{Jing-Fei Zhang\orcidlink{0000-0002-3512-2804}}
		\affiliation{Liaoning Key Laboratory of Cosmology and Astrophysics, College of Sciences, Northeastern University, Shenyang 110819, China}
		
		\author{Xin Zhang\orcidlink{0000-0002-6029-1933}}\email{zhangxin@neu.edu.cn}
		\affiliation{Liaoning Key Laboratory of Cosmology and Astrophysics, College of Sciences, Northeastern University, Shenyang 110819, China}
		\affiliation{MOE Key Laboratory of Data Analytics and Optimization for Smart Industry, Northeastern University, Shenyang 110819, China}
		\affiliation{National Frontiers Science Center for Industrial Intelligence and Systems Optimization, Northeastern University, Shenyang 110819, China}

		\begin{abstract}
			Traditional diagnostics of black hole phase transitions rely on thermodynamic quantities defined at the event horizon or asymptotic boundary. Here, we demonstrate that the near-singularity geometry offers a sharp, independent probe of both first-order phase transitions and supercritical crossover. For scalarized AdS black holes exhibiting a first-order phase transition, the Kasner exponent $p_t$, which characterizes the approach to the singularity, undergoes a dramatic transformation. On one side of the transition, $p_t$ oscillates strongly with temperature, reflecting violent interior dynamics. On the other side, it becomes a smooth, monotonically varying function. These two distinct behaviors converge as the critical point is approached. Beyond the critical point, in the supercritical region,  $p_t(T)$ develops a distinct extremum, defining a ``Kasner crossover line'' that is entirely independent of traditional thermodynamic (Widom line) or dynamic (Frenkel line) criteria. Our work establishes the near geometry of singularity of scalarized black hole as a novel class of diagnostics for phase transitions, revealing that a change in the macroscopic thermodynamic state fundamentally reshapes the deepest interior structure of spacetime.
		\end{abstract}

		\maketitle
		\section{Introduction}
		The thermodynamics of black holes, rooted in the physics of the event horizon \cite{Hawking:1975vcx,Bekenstein:1973ur,Hawking:1982dh}, has revealed a rich phase structure akin to that of ordinary matter, including first-order phase transitions and critical phenomena \cite{Chamblin,Kubiznak2012,Ruppeiner,Wei20152016,Wei2019,Wei2024}. In the Anti-de Sitter (AdS) context, these transitions are particularly intriguing due to their holographic dual interpretation \cite{Maldacena:1997re,Hartnoll:2008vx,Hartnoll:2008kx,Cai:2010cv,Cai:2010zm,Li:2011xja,Cai:2013aca,Cai:2015cya}. However, all existing probes—such as free energy, heat capacity, or order parameters—are defined on the outside of black holes. A fundamental question remains: does a macroscopic change in the thermodynamic state of a black hole leave any imprint on the geometry inside the horizon, deep in the region leading to the singularity?

		Recent studies of the interior of scalarized black holes have shown that the region beyond the horizon is not static. Instead, it undergoes a series of dynamical epochs—the collapse of the Einstein-Rosen bridge and Josephson oscillations—before ultimately settling into a universal, homogeneous Kasner epoch \cite{Hartnoll:2020fhc,Cai:2020wrp,Wang:2020nkd,Liu:2021hap,An:2021plu,Mansoori:2021wxf,Liu:2021hap,Cai:2021obq,An:2022lvo,Sword:2022oyg,Liu:2022rsy,Xu:2023fad,Gao:2023zbd,Gao:2023rqc,Zhang:2025hkb,Zhang:2025tsa,Xu:2025edz,Gao:2025qqa,Xiong:2026npi}. This final Kasner geometry is fully characterized by a set of exponents, such as $p_t$, which encode the behavior of spacetime and matter fields as they approach the singularity. This universality suggests a compelling possibility: the interior Kasner geometry might serve as a ``record'' of the black hole's macroscopic history.

		In this paper, we demonstrate that this is indeed the case. By studying scalarized AdS black holes that undergo a first-order phase transition, we show that the Kasner exponent $p_t$ is an exquisitely sensitive probe of the transition. We find that $p_t(T)$ exhibits dramatically different behaviors on the two stable branches of the transition—oscillatory on one and smooth on the other—which converge at the critical point. Furthermore, in the supercritical region, the temperature dependence of $p_t$ develops a distinct extremum, defining a novel ``Kasner crossover line'' that is independent of traditional thermodynamic or dynamical criteria. Our findings reveal that the influence of a black hole's phase transition is not confined to the exterior; it fundamentally reshapes the geometry all the way down to the singularity itself. It is worth noting that the cases discussed here are all black holes after scalarization. In the original asymptotically flat black holes without scalar hair, no phase transitions exist.
		
		
		This work marks the first identification of the black hole singularity as a new class of probe for phase transitions. By offering a direct and sensitive window into the interior dynamics during a first-order phase transition, it uncovers a deep physical insight: the alteration of a black hole's macroscopic thermodynamic state is not confined to the exterior; its influence penetrates the horizon and fundamentally reconstructs the geometry at the most fundamental level—the singularity itself. The remainder of this paper is organized as follows. In Section \ref{section2}, we present the model we use, the complete equations of motion, and the formula for the free energy. In Section \ref{section3}, we discuss the relation between the phase structure and the interior geometry of the black hole. Finally, we provide a summary in Section \ref{section4}.

		\section{Scalarized black hole with charged scalar field}\label{section2}
		We begin with a simple action in the 3+1-dimensional spacetime, considering a charged scalar field $\psi$ along with a nonlinear term $\lambda(\psi^{\ast}\psi)^{2}$. This term is introduced primarily to induce a first-order phase transition in the system. The total action takes the following form
		\begin{align}
			&S=\frac{1}{16\pi G}\int d^{4}x\sqrt{-g}(R-2\Lambda+\mathcal{L}_{m}),\label{Lagg}\\
			&\mathcal{L}_{m}=-\frac{1}{4}F_{\mu\nu}F^{\mu\nu}-D_{\mu}\psi^{\ast}D^{\mu}\psi-\frac{m^{2}}{L^2}\psi^{\ast}\psi-\lambda(\psi^{\ast}\psi)^{2}.
		\end{align}
		In this setup, $\Lambda = -3/L^2$, where $L$ is the AdS radius \cite{Dolan:2010ha}. In the extended phase space,  the parameter $L$ is related to the pressure via $P=3/(8\pi L^2)$ \cite{Kubiznak:2012wp}. Here, $F_{\mu\nu}=\nabla_{\mu}A_{\nu}-\nabla_{\nu}A_{\mu}$ is the Maxwell field strength and $D_{\mu}\psi=\nabla_{\mu}\psi-iq A_\mu\psi$ is the standard covariant derivative term of the charged scalar field $\psi$. For simplicity and without loss of generality, we set $G= 1$.
		
		The equations of motion for the matter fields are obtained by varying the action with respect to $\psi$ and $A_{\mu}$, respectively. We adopt an ansatz of the form
		\begin{align}
			\psi=\psi(z),~A_\mu dx^\mu=\phi(z)dt.
		\end{align}
		The Einstein equation is
		\begin{align}
			R_{\mu\nu}-\frac{1}{2}(R-2\Lambda)g_{\mu\nu}=\frac{1}{2}\mathcal{T}_{\mu\nu},
		\end{align}
		where $\mathcal{T}_{\mu\nu}$ is the stress-energy tensor of the matter fields
		\begin{align}
			\mathcal{T}_{\mu\nu}=&(-\frac{1}{4}F_{\alpha\beta}F^{\alpha\beta}
			-D_{\alpha}\psi^{\ast}D^{\alpha}\psi-\frac{m^{2}}{L^2}\psi^{\ast}\psi-\lambda(\psi^{\ast}\psi)^{2})g_{\mu\nu}\nonumber\\
			&+(D_{\mu}\psi^{\ast}D_{\nu}\psi+D_{\nu}\psi^{\ast}D_{\mu}\psi)+F_{\mu\alpha}F^{\alpha}_{\nu}.
		\end{align}
		The metric of the black hole is given by
		\begin{align}
			ds^{2}=\frac{1}{z^2}(-f(z)e^{-\chi(z)}dt^{2}+\frac{1}{f(z)}dz^{2}+dx^{2}+dy^{2}),
		\end{align}
		In this case, the temperature of the black hole is defined as
		\begin{align}
			T=\frac{1}{4\pi}f'(z_h)e^{-\chi(z_h)/2}.
		\end{align}
		The complete set of equations of motion are 
		\begin{align}
			0=&z^2e^{-\chi/2}(e^{\chi/2}\phi')'-\frac{2q^2\psi^2}{f}\phi,\label{EM1}\\
			0=&z^2e^{\chi/2}(\frac{e^{-\chi/2}f}{z^2}\psi')'-
			(\frac{m^2}{L^2z^2}\psi-\frac{q^2e^{\chi}\phi^2}{f}\psi+\frac{2\lambda}{z^2}\psi^3),\label{EM2}\\
			0=&\frac{\chi'}{z}-(\frac{q^2e^{\chi}}{f^2}\phi^2\psi^2+\psi'^2),\label{EM3}\\
			0=&4e^{\chi/2}z^4(\frac{e^{-\chi/2}f}{z^3})'-(\frac{2m^2}{L^2}\psi^2+z^4e^{\chi}\phi'^2-\frac{12}{L^2}+2\lambda\psi^4).\label{EM4}
		\end{align}
		in which
		\begin{align}
			f(z)=1-2z^3M(z).
		\end{align} 
		
		To numerically solve the equations of motion, we need to provide the expansions at the horizon $z\rightarrow z_h$ and at infinity $z\rightarrow 0$. The expansions at the horizon are
		\begin{align}
			\phi(z)=&\phi_{h_1}(z-z_h)+\phi_{h_2}(z-z_h)^2+\cdots,\label{horizonExpPhi}\\
			\psi(z)=&\psi_{h_0}+\psi_{h_1}(z-z_h)+\cdots,\label{horizonExpPsi}\\
			\chi(z)=&\chi_{h_0}+\chi_{h_1}(z-z_h)+\cdots,\label{horizonExpChi}\\
			M(z)=&\frac{1}{2z^3_h}+M_{h_1}(z-z_h)+\cdots.\label{horizonExpM}
		\end{align}
		Near AdS boundary, the expansion of
		the functions are
		\begin{align}
			\phi(z)&=\mu-z\rho+\cdots,\\
			\psi(z)&=z\psi^{(1)}+z^2\psi^{(2)}+\cdots,\\
			\chi(z)&=\chi_{b_0}+z^3\chi_{b_3}+\cdots,\\
			M(z)&=M_{b_0}+zM_{b_1}+\cdots.
		\end{align}
		In the AdS expansion of $\phi(z)$, $\mu$ can be regarded as the chemical potential of the dual field theory.
		In this paper, we work in the canonical ensemble by fixing the total charge density $\rho$. The free energy of the black hole system is given by the on-shell action multiplied by the temperature which reads
		\begin{align}
			\widetilde{F}=-T ln\mathcal{Z}, \mathcal{Z}=e^{-S^E},
		\end{align}
		For simplicity and without loss of generality, we neglect the spatial volume, thereby yielding the following form for the Helmholtz free energy
		\begin{align}
			F=\frac{2\kappa_g^2}{V_2}\widetilde{F},
		\end{align}
		in which
		\begin{align}
			S_E=&-\frac{1}{16\pi G}\int d^{4}x\sqrt{-g}(R-2\Lambda+\mathcal{L}_{m})\nonumber\\
			&+\frac{1}{8\pi G}\int d^{3}x\sqrt{-h}\left(\frac{1}{2}n_aF^{ab}A_b+K+\frac{2}{L}\right).\label{EuclideaS}
		\end{align}
		Here $\mathcal{L}_{m}$ is the Lagrangian density, and the specific form of $K$ is $K_{\mu\nu}=-h_\mu^{~\rho}\nabla_\rho n_\nu$, where $n_a$ is the unit normal on the boundary surface (see, e.g., Refs. \cite{Cai:2002mr,Nie:2014qma}) and $\kappa_g^2=8\pi G$. The final expression for the Helmholtz free energy can be written as
		\begin{align}
			F&=\mu\rho-2M_{b0},\label{finalFreeEnergy}
		\end{align}
		where the term $\mu\rho$ originates from the integral of the surface term $\frac{1}{2}n_aF^{ab}A_b$. Substituting the asymptotic expansion of $\phi(z)$ at the AdS boundary yields the above form. The term $2M_{b0}$ mainly comes from the contribution of the metric.
		
		In the remainder of this paper, we will present the specific results for the free energy. For convenience, we fix $q = 1$ and $z_h = 1$ in the following. Numerically, we solve the complete equations of motion using the shooting method. In this work, we have only considered global stability, because we only need to find the most stable scalarized black hole solution and do not need to consider local stability.

		\section{Interior solutions of black holes and Kasner geometry}\label{section3}
		The interior solution of a black hole can be derived by combining numerical results with analytical methods, while neglecting infinitesimal quantities in certain field equations inside the black hole. Through analytical treatment and setting the scalar field as $\psi=\sqrt{2}\int(\Psi/z)dz$, where $\Psi=\alpha$ (see Ref.~\cite{Hartnoll:2020fhc} for details), the interior solution can be expressed in the following forms
		\begin{align}
			f=-f_Kz^{3+\alpha^2},~\psi=\alpha\sqrt{2}\log z,~\chi=2\alpha^2\log z+\chi_K,\label{SolutionSimpleEq}
		\end{align}
		where $f_K$ and $\chi_K$ are constants. Using the solutions (\ref{SolutionSimpleEq}), we can obtain the metric components $g_{tt}$ and $g_{zz}$ in the following forms
		\begin{align}
			g_{tt}=f_Kz^{1-\alpha^2},~g_{zz}=-\frac{1}{f_K}z^{-5-\alpha^2}.
		\end{align}
		
		Transforming the $z$ coordinate to the proper time $\tau\sim z^{-3/2-\alpha^2/2}$, we can obtain the Kasner form of the metric
		\begin{align}
			ds^2&=-d\tau^2+c_t\tau^{2p_t}dt^2+c_x\tau^{2p_x}(dx^2+dy^2),\nonumber\\
			\psi&=-p_{\psi}\log\tau,
		\end{align}
		in which $c_t$ and $c_x$ are constants and
		\begin{align}
			p_t=\frac{\alpha^2-1}{3+\alpha^2},~~p_x=\frac{2}{3+\alpha^2},~~p_{\psi}=\frac{2\sqrt{2}\alpha}{3+\alpha^2}.\label{KasnerMtric}
		\end{align}
		Here, the Kasner metric is merely used to describe the geometry near the singularity inside the black hole horizon.
		After crossing the black hole horizon, the scalar field transitions through several dynamical epochs: the collapse of the Einstein-Rosen bridge, Josephson oscillations, and a stable Kasner epoch (or reaches a stable Kasner epoch after a Kasner inversion) \cite{Hartnoll:2020fhc}. In the Kasner epoch, the scalar field stably approaches the singularity, and information at this stage can be extracted via the Kasner exponent $p_t$. Notably, the relationship between 
		$p_t$ and temperature exhibits a well-defined inverse periodic behavior near the critical point. In Ref.~\cite{Zhao:2025odj} the authors investigated the properties of this inverse periodic behavior in detail and showed that it is not broken by higher-order nonlinear terms, but merely stretched or compressed. While such inverse periodic or oscillatory behavior carries limited physical significance in conventional second-order phase transitions, the situation becomes entirely different when the system undergoes a first-order phase transition.
		
		\begin{figure}[!htbp]
			\center
			\includegraphics[width=0.9\columnwidth]{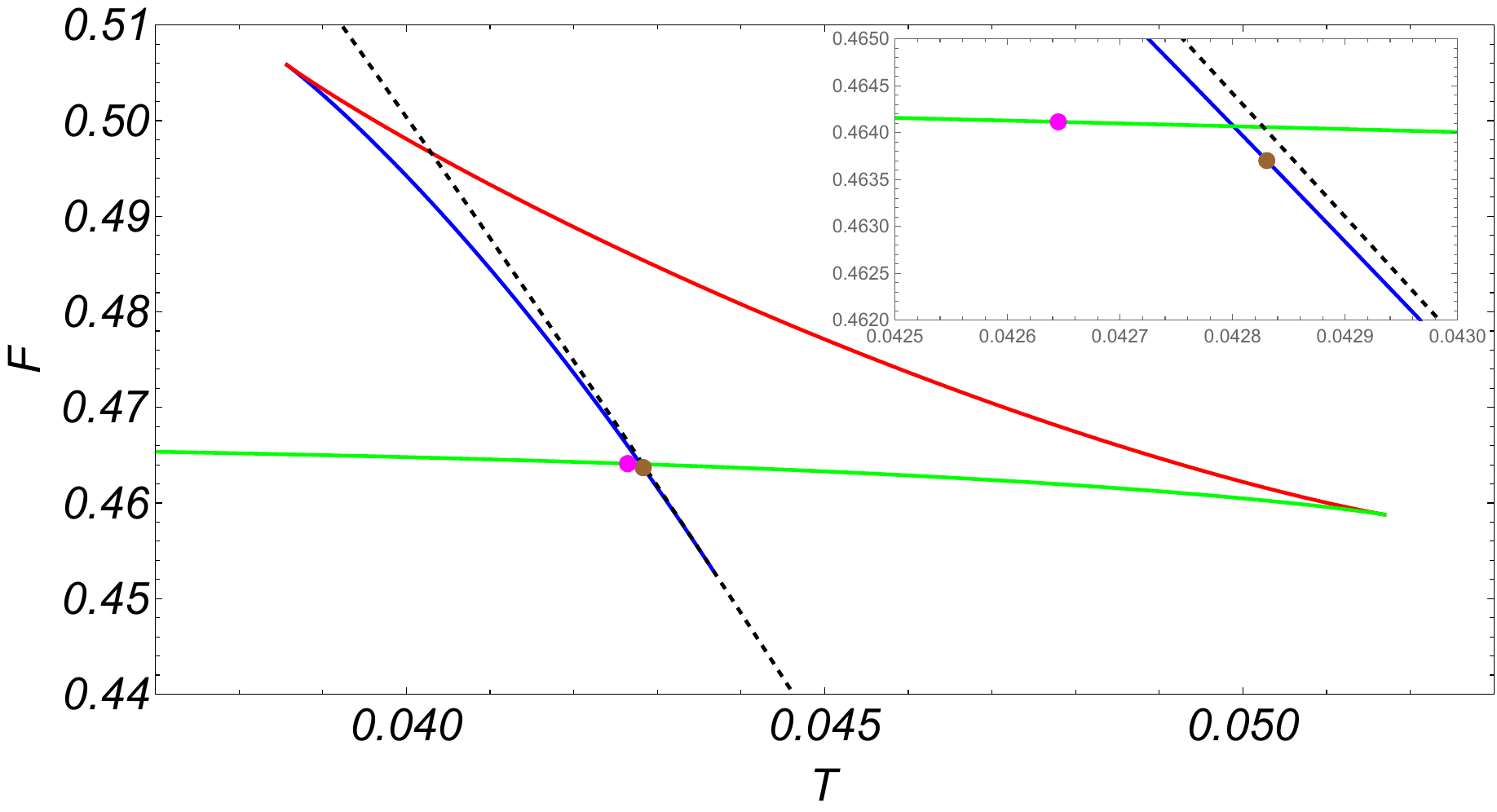}
			\includegraphics[width=0.9\columnwidth]{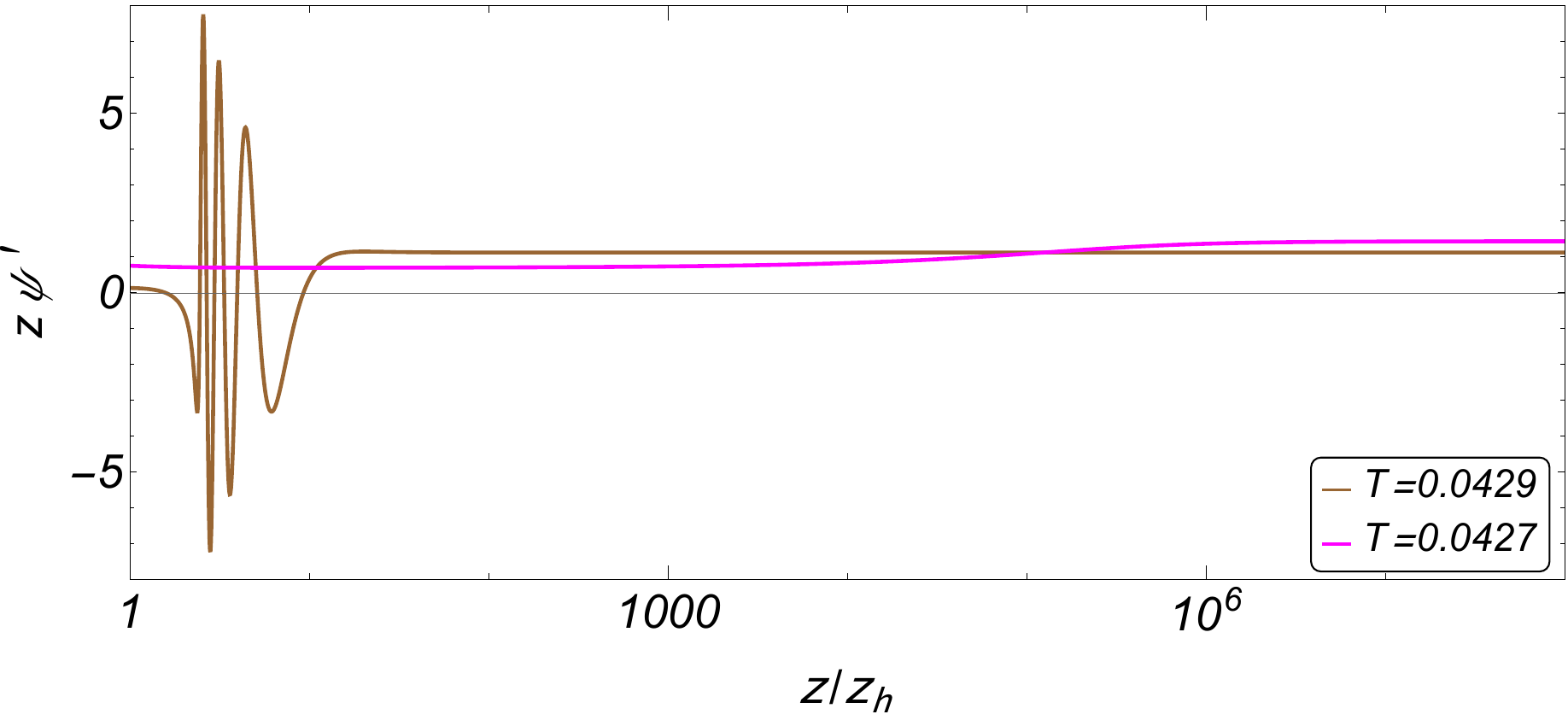}
			\caption{The free energy of first-order phase transition and interior structure of the scalar field with $\lambda=-0.3$, $P=0.03~(L=1.994)$. The top panel is free energy of the first-order phase transition. The dashed line represents the normal solution. The blue and green solid lines correspond to scalarized black hole solutions 1 and 2, respectively. The red solid  line indicates the unstable black hole solution. The bottom panel is the behavior of the scalar field inside the black hole. Curves of different colors correspond to selected points on the respective solutions in the top panel.}\label{firstOrderCURVE}
		\end{figure}

		A sufficiently negative coupling coefficient $\lambda$ triggers a first-order phase transition in our model \cite{Zhao:2025tqq,Zhao:2025vtr}. The upper panel of Fig.~\ref{firstOrderCURVE} shows the characteristic ``swallowtail'' behavior of the free energy $G(T)$, indicating a first-order phase transition between two stable scalarized black hole branches (blue and green curves). To understand how this thermodynamic discontinuity affects the deep interior, we examine the scalar field profile $\psi(z)$ from the horizon $(z=z_h)$ to the singularity $(z\rightarrow\infty)$, as shown in the lower panel. The field on the high-temperature branch (blue curve in upper panel, exemplified by the brown line in lower panel) exhibits highly oscillatory behavior as it approaches the singularity. In stark contrast, the field on the low-temperature stable branch (green curve in upper panel, magenta line in lower panel) decays smoothly and monotonically, leading to a qualitatively different near-singularity geometry.

\begin{figure}[!htbp]
			\center
			\includegraphics[width=0.9\columnwidth]{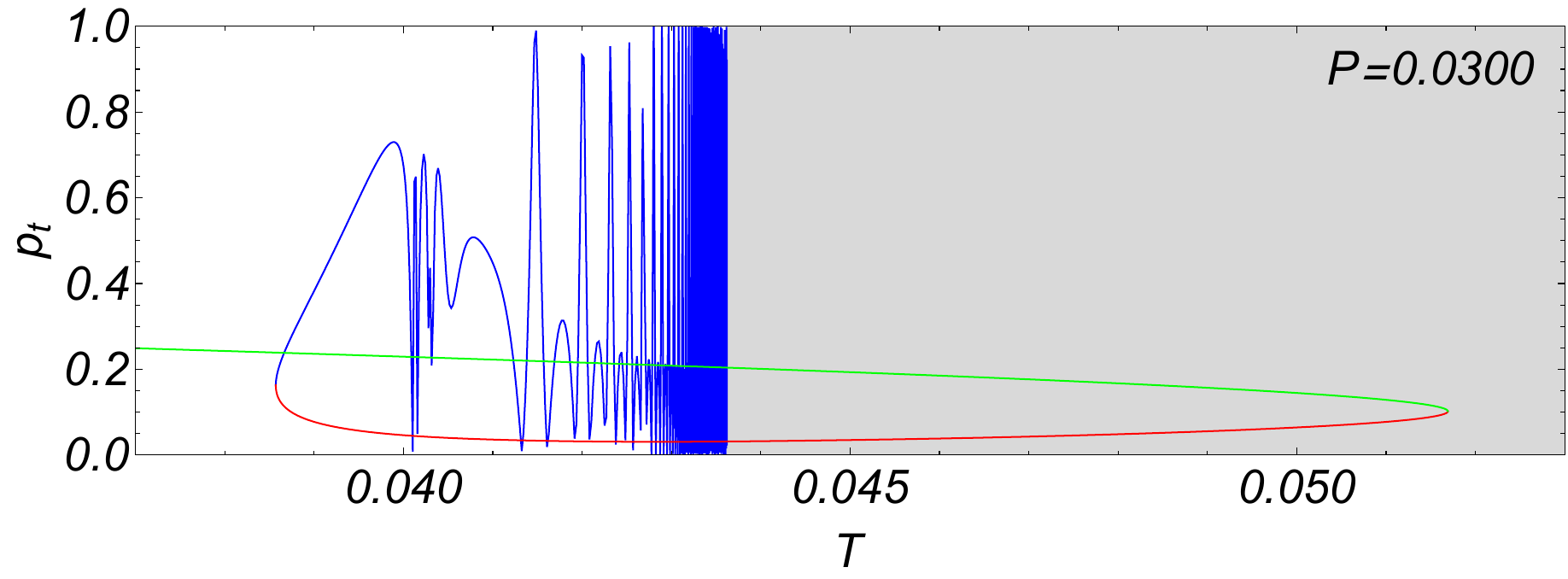}
			\includegraphics[width=0.9\columnwidth]{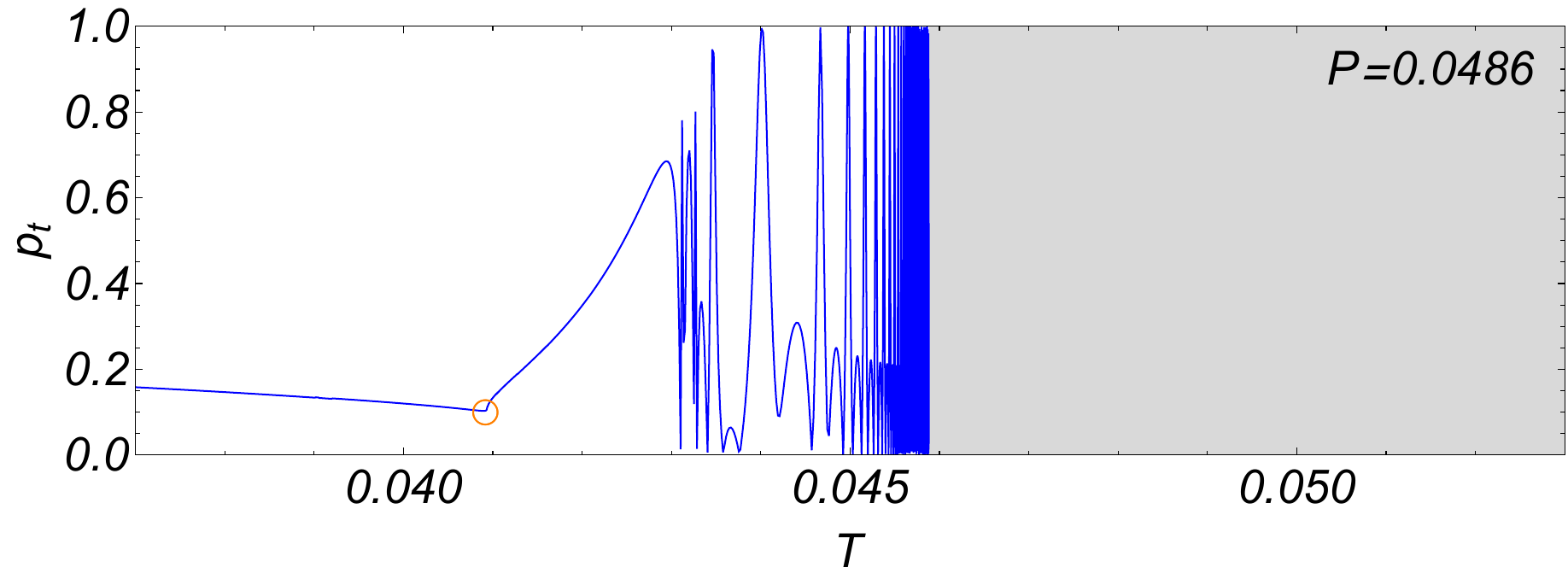}
			\caption{The dependence of the kasner exponents $p_t$ and a function of temperature $T$. The top panel shows $p_t(T)$ in the first-order phase transition region, and the bottom panel shows $p_t(T)$ in the supercritical region.}\label{interiorptT}
		\end{figure}

		Apart from the distinct differences in the behavior of the interior of a black hole before and after the phase transition, significant differences can also be observed in the Kasner exponents $p_t$. We computed the variation of Kasner exponents $p_t$ with temperature $T$ and present the computation results in the top panel of Fig.~\ref{interiorptT}. For scalarized black hole solution 1 near the critical point, its Kasner exponents $p_t$ exhibit strong oscillatory behavior with changing temperature. In contrast, scalarized black hole solution 2 demonstrates a smooth and stable profile.

		\begin{figure}[!htbp]
			\center
			\includegraphics[width=1\columnwidth]{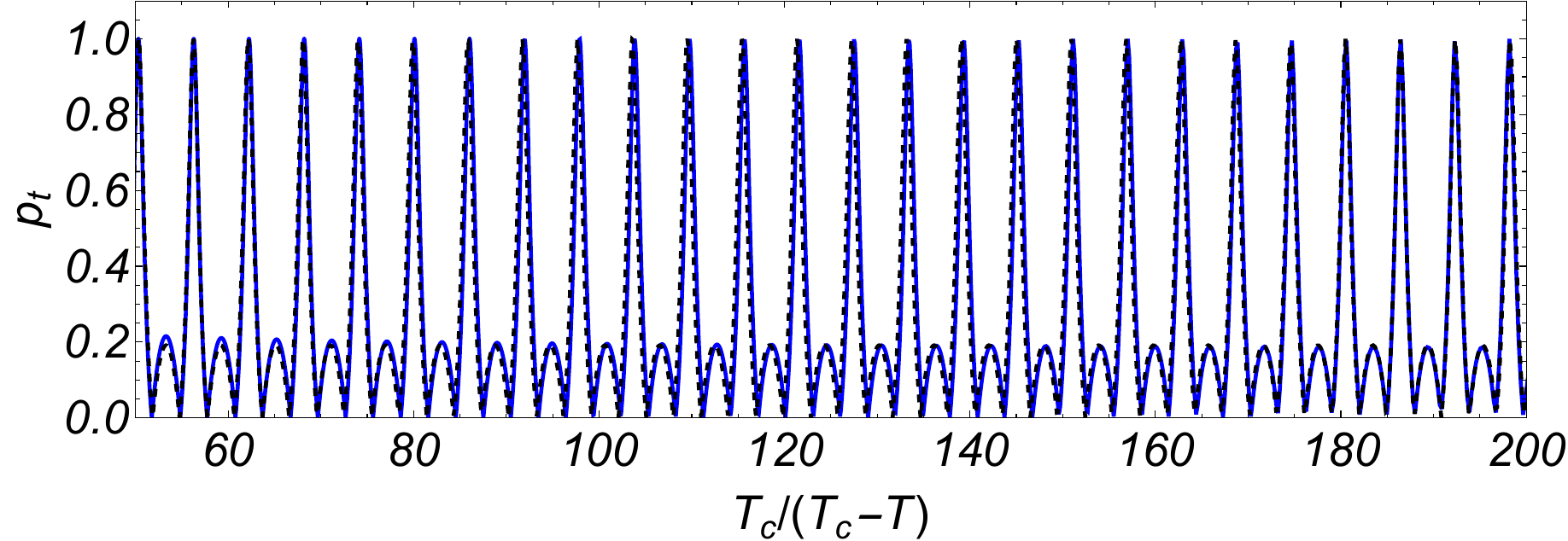}
			\caption{The relationship between the Kasner exponent $p_t$ and $T_c/(T_c-T)$ for the first-order phase transition. The blue solid lines represents the original $p_t$ data, and the black dashed line denote the data derived from $p_t(\alpha_0)$. The parameter values are $A=1.55,B=0.53,C=1.54$}\label{ptInverse}
		\end{figure}
		
		We also compute the inverse periodic behavior of $p_t$ near the critical point when the black hole undergoes a first-order phase transition. As shown in Ref.~\cite{Hartnoll:2020fhc}, the inverse periodic behavior persists even in the presence of a first-order phase transition. As already described in Ref.~\cite{Hartnoll:2020fhc}, this behavior can be described by the analytical formula $p_t(\alpha_0)$, in which $\alpha_0 = -\sqrt{8}\frac{A}{\pi}\sin\left(\frac{B}{1-T/T_c} + C\right).$ We present the results in Fig.~\ref{ptInverse}.

		\begin{figure}[!t]
			\center
			\includegraphics[width=1\columnwidth]{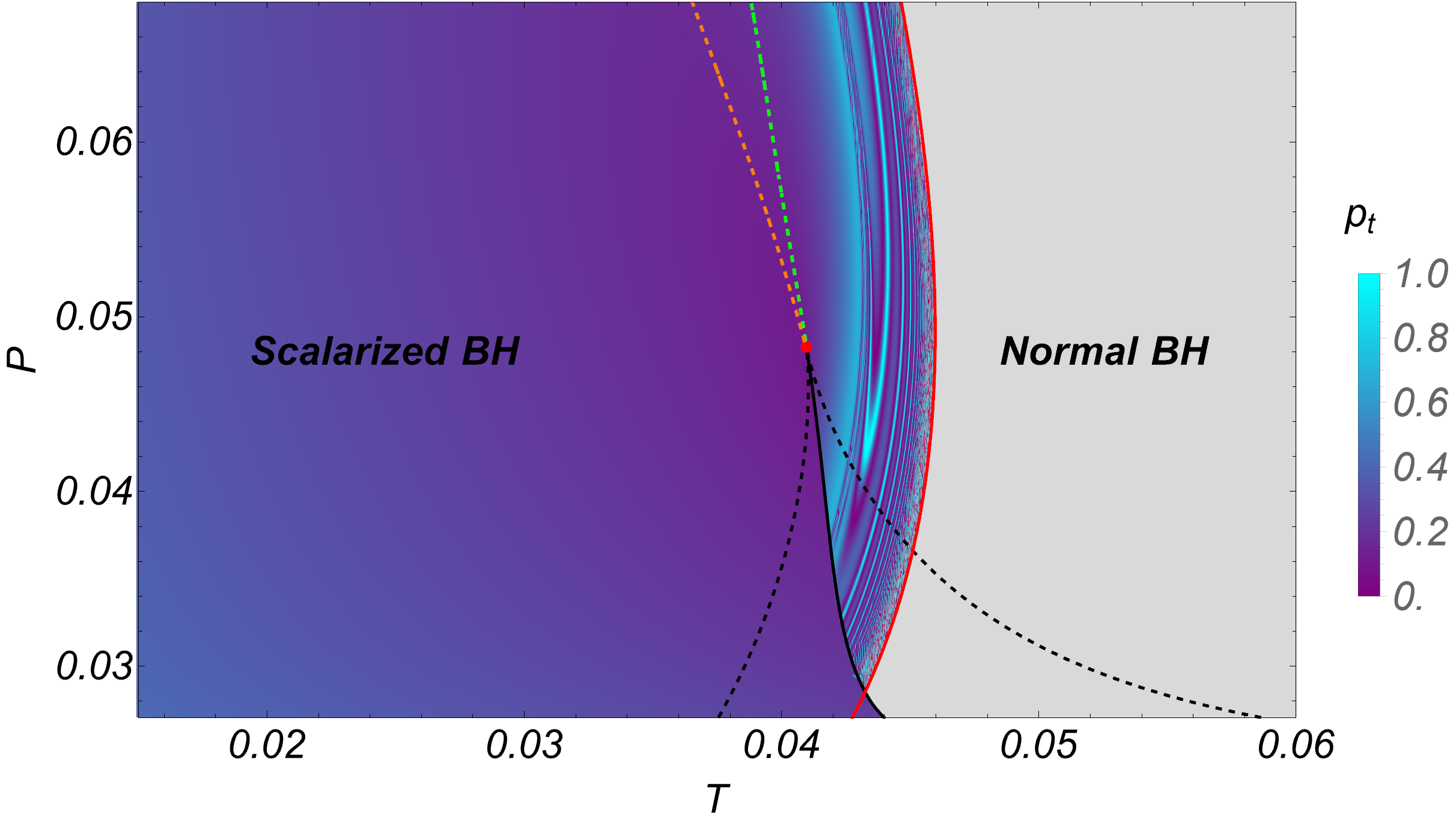}
			\caption{The phase diagram of the Kasner exponent $p_t$ as a function of temperature and pressure. The red line indicates the critical point from the normal phase to the scalarized phase. The black solid line denotes the first-order phase transition points, while the black dashed lines represent the spinodal lines of the first-order phase transition. The orange dashed line mark the points where the slope of $p_t$ changes abruptly, and the green dashed line corresponds to the Widom line, defined as $(\partial^2 G/\partial T^2)/(\partial G/\partial T)$.}\label{PTphaseDiagram}
		\end{figure}



		Generally, as the condensate (the value at the horizon) increases, three distinct stages emerge. First, when the scalar field is small, $p_t$ exhibits regular inverse period oscillations with a stable period. If this behavior continues beyond the turning point of the condensate curve, the period would invert. From the perspective of period stability, we argue that this regular inverse period oscillation cannot extend beyond the first inflection point encountered after the critical point. After the first stage of stable inverse period oscillation ends, the behavior of $p_t$ enters a second stage, characterized by irregular oscillations compared to the first stage. After a small increase in the condensate, these irregular oscillations give way to a third stage, where $p_t$ no longer oscillates with increasing condensate. Due to the presence of an unstable branch between the two stable branches during a first-order phase transition, the irregular oscillations of the second stage are likely to be consumed within this unstable branch, thus failing to reach the stable branch with a higher condensate. Therefore, in general, $p_t$ exhibits a sudden transition from oscillatory to non-oscillatory behavior across a first-order phase transition.

		Now let us discuss the reason for this phenomenon. For the interior solution of the black hole, the general procedure is as follows. We first solve the field equations outside the black hole in the interval $z \in [0, z_h]$. Once the specific solutions of Eqs.~(\ref{EM1}-\ref{EM4}) are obtained, we know the value of $\psi$ at the horizon, i.e., the magnitude of the scalar condensate and the numerical values of the other field equations. The corresponding free energy can then be obtained via Eq.~(\ref{finalFreeEnergy}). After these values are obtained, we extend the solution interval to $[z_h, z_s]$, where $z_s$ is chosen such that the interior solution evolves until a stable Kasner plateau is reached (the value of this plateau can be converted into the Kasner exponent $p_t$ through Eq.~(\ref{KasnerMtric})). This process of solving the interior solution can be viewed as a time evolution of the field equations along the $z$ direction from $z_h$ toward the singularity at infinity, where $z$ plays the role of time and the values at the horizon serve as the initial conditions for the evolution toward the singularity. In other words, the behavior of the black hole interior is completely determined by the set of values $\{\psi(z_h), \phi(z_h), \chi(z_h), M(z_h)\}$. For small $\psi(z_h)$, the interior solution undergoes violent Josephson oscillations (the brown curve in Fig.~\ref{firstOrderCURVE}). In contrast, for large $\psi(z_h)$, the behavior is relatively smooth (the magenta curve in Fig.~\ref{firstOrderCURVE}). After scalar condensation occurs, $\psi(z_h)$ increases gradually as the temperature decreases. During this process, the interior solution gradually consumes the Josephson oscillation region of the system. If this is converted into the Kasner exponent $p_t$, this inevitably leads to the situation shown in Fig.~\ref{interiorptT}, where the blue solution corresponding to the small condensate region exhibits highly oscillatory behavior in $p_t$, while the green solution corresponding to the large condensate region exhibits smooth behavior.

		
		Finally, in Fig.~\ref{PTphaseDiagram} we present a density plot of the Kasner exponent $p_t$ as a function of temperature and pressure. In the high temperature region, there are no scalarized black hole solutions exist. As the temperature decreases, hairless black holes undergo a second-order phase transition to scalarized black holes. In the low-pressure region, due to the presence of a first-order phase transition, the solid black line denoting the phase transition points delineates two distinct regions: the near-critical point scalarized black hole 1 region, where $p_t$ exhibits highly oscillatory behavior, and the scalarized black hole 2 region, where $p_t$ varies more gradually. As pressure increases, the behavior of $p_t$ near the phase transition point gradually becomes similar. This feature resembles that of conventional gas-liquid phase transitions, where the gas and liquid phases exhibit similar dynamic and thermodynamic properties in the high-temperature and high-pressure subcritical region. The authors in Ref.~\cite{Zhao:2025vtr} have discussed in detail the role of pressure in regulating scalarized black holes.

		When the pressure exceeds the critical point, the system enters the supercritical region. As established in previous studies, the supercritical region can be further subdivided by thermodynamic criteria \cite{Zhao:2024jhs,Xu:2025jrk,Zhao:2025ecg,Wang:2025ctk,Li:2025lrq,Guo:2026xlk} (e.g., the Widom line) or dynamic criteria \cite{Zhao:2024jhs,Zhao:2025ecg,Li:2025tdd} (e.g., the Frenkel line).  However, when discussing supercritical black holes, we typically focus on external information which is derived from thermodynamics or dynamics. In this work, we find that information from the black hole singularity also provides a criterion for the supercritical crossover that is completely independent of thermodynamics and dynamics. 
		The bottom panel of Fig.~\ref{interiorptT} illustrates the variation of $p_t$ with temperature at the critical point, where an obvious abrupt change point, marked by an orange circle, separates the highly oscillatory region from the region with a smooth variation in curvature. In Fig.~\ref{PTphaseDiagram}, we denote these supercritical crossover points with an orange dashed line, called the Kasner crossover line.
		
		Similar to the changes occurring inside the black hole before and after the first-order phase transition discussed earlier, the interior changes of the black hole in the supercritical region are even more significant. Although traditional thermodynamic and dynamical methods have provided criteria for the supercritical crossover, as reflected in the central theme of this work, none of these criteria focus on the interior information of the black hole. Our results show that the interior of the black hole can also distinguish supercritical subphases in the supercritical region, and this distinction is made from a completely new perspective.
		
		\section{Conclusion}\label{section4}
		In this paper, we reveal for the first time that the near-singularity geometry of a scalarized black hole can serve as a sensitive probe of first-order phase transitions. Through a study of scalarized AdS black holes undergoing a first-order phase transition, we find that the Kasner exponent $p_t$, which characterizes the behavior near the singularity, exhibits markedly distinct features on either side of the transition: on one branch (scalarized branch 1), $p_t$ oscillates strongly with temperature, while on the other branch (scalarized branch 2), it displays a smooth and stable profile. As pressure and temperature approach the critical point, these two behaviors gradually converge, consistent with the physical picture of phase indistinguishability near criticality in conventional fluids.
		
		More importantly, when the system crosses the critical point into the supercritical region, a distinct extremum emerges in the temperature dependence of $p_t$. This finding provides a novel criterion—completely independent of traditional thermodynamic (e.g., Widom line) or dynamic (e.g., Frenkel line) criteria—for characterizing the supercritical crossover behavior in black hole systems. We term this new delineation the ``Kasner crossover line''.
		
		Our work establishes the singularity of scarized black hole as a new class of diagnostic tool for first-order phase transitions. It not only offers a sharp, interior observational window into first-order phase transitions but also reveals a profound physical picture: a change in the macroscopic thermodynamic state of a black hole penetrates the event horizon and may fundamentally reshapes the deepest structure of spacetime. This finding extends our understanding of black hole phase transitions from external thermodynamic quantities to the interior spacetime geometry, providing a novel perspective for studying black hole phase transitions.
		
		There remains an open question. Our current theory only considers this particular model of scalarized black holes. Whether the same conclusions hold for other types of scalarized black holes, such as those in Einstein-Maxwell-dilaton models, Born-Infeld gravity, or Gauss-Bonnet gravity, is a question worth investigating. We plan to study these issues in detail in the future.

		\section*{Acknowledgements}
		This work is partially supported by the National Natural Science Foundation of China (Grant Nos. 12533001, 12575049, 12473001, 12205039, 12305058, 11965013 and 12575054). ZYN is partially supported by Yunnan High-level Talent Training Support Plan Young $\&$ Elite Talents Project (Grant No. YNWR-QNBJ-2018-181). The work is also supported by the National SKA Program of China (grant Nos. 2022SKA0110200
		and 2022SKA0110203) and the 111 Project (Grant No. B16009).

		\bibliographystyle{apsrev4-1}
		\bibliography{inbh}
		
		

	\end{document}